\documentstyle[12pt]{article}                 %% AMS-LaTeX
\newlength{\dinwidth}
\newlength{\dinmargin}
\newcommand{\RR}{{\mbox{\bf R}}}
\newcommand{\bx}{{\mbox{\boldmath $x$}}}
\newcommand{\bp}{{\mbox{\boldmath $p$}}}
\newcommand{\sbx}{{\mbox{\footnotesize \boldmath $x$}}}
\newcommand{\sbp}{{\mbox{\footnotesize \boldmath $p$}}}

\newcommand{\Wbo}{{W^{(0)}_\beta (x, m)}}
\newcommand{\Dxm}{{\Delta (x, m)}}
\newcommand{\hpx}{{h^{(+)}({\bx}, m)}}
\newcommand{\hmx}{{h^{(-)}({\bx}, m)}}
\newcommand{\hpmx}{{h^{( \pm )}({\bx}, m)}}
\newcommand{\intf}{{\int d x_0 \, f ( x_0 )}}

\newcommand{\Tr}{{\mbox{Tr}}}
\newcommand{\pde}{{{\partial}_{0}}}

\newcommand{\Dpx}{{D^{(+)}_\beta ({\bx}, m)}}   
\newcommand{\Cpx}{{C^{(+)}_\beta ({\bx}, m)}}   
\newcommand{\Cplx}{{C^{(+)}_\beta ({\lambda \bx}, \lambda^{-1} m)}}  
\newcommand{\Dpp}{{\widetilde{D}^{(+)}_\beta ({\bp}, m)}}       
\newcommand{\Dmx}{{D^{(-)}_\beta ({\bx}, m)}}   
       
\newcommand{\Dpmx}{{D^{( \pm )}_\beta ({\bx}, m)}}      
  
\newcommand{\Rpx}{{R^{(+)}_\beta ({\bx}, m)}}   
\newcommand{\Rpp}{{\widetilde{R}^{(+)}_\beta ({\bp}, m)}}       

\begin{document}
\title{The Unmasking of Thermal Goldstone Bosons}
\author{Jacques Bros$^a$ \,  and  \, Detlev Buchholz$^b$
\\[5mm]
${}^a \,$ Service de Physique Th\'eorique, CEA--Saclay\\
F--91191 Gif--sur--Yvette, France\\[2mm]
${}^b \,$ Institut f\"ur Theoretische Physik,
Universit\"at G\"ottingen\\
D--37073 G\"ottingen, Germany}
\date{}
\maketitle
%
%%%%%%%%%%%%%%%          TEXT DES ARTIKELS          %%%%%%%%%%%%%%%%%%%%
%

\begin{abstract} 
\noindent The problem of extracting the modes of Goldstone bosons  
from a thermal background is reconsidered in the framework of 
relativistic quantum field theory. It is shown that in the case of 
spontaneous breakdown of an internal bosonic symmetry 
a recently established decomposition of 
thermal correlation functions contains certain specific contributions 
which can be attributed to a particle of zero mass. \\[2mm]
PACS-numbers: 11.10.Cd, 11.10.Wx, 11.30.Qc 
\end{abstract}

\noindent The relation between the spontaneous breakdown of symmetries
and the structure of the energy--momentum spectrum is known to be more
subtle for thermal states than in the case of the vacuum. 
According to the thermal version of the Goldstone theorem, there appears
a discrete zero energy mode in current--field correlation functions
whenever the symmetry corresponding to the current is spontaneously
broken, see for example \cite{Ka}. But, 
in contrast to the vacuum case, there need not hold a
sharp energy--momentum dispersion law which would allow one to identify
clearly the Goldstone particles in the usual way. As a matter of 
fact, the existence of such a law would imply
that there is no interaction \cite{NaReTh}. 

The fact that there appear in general only low--energy artifacts of
Goldstone particles in thermal correlation functions can easily be
understood in heuristic physical terms: the particles collide
with large probability with other constituents of the state and
thereby change their energy and momentum. Hence discrete 
($\delta$--function) contributions
in the correlation functions corresponding to events where a particle
excitation remains unaffected by the thermal background are dynamically
suppressed. It is only at zero energy where such 
contributions can survive since soft massless particles do not
participate in any reactions according to well--known low energy
theorems (cf.\ \cite{Um} and references quoted therein). 

There appears yet another complication in the case of spontaneously broken 
geometric symmetries, such as Lorentz transformations, and supersymmetries.
As was first pointed out in \cite{Oj}, the discrete zero energy modes are in 
general not affiliated with a particle in these cases but are due to 
particle--hole pairs. In view of these facts the foundations for a 
particle interpretation of the Goldstone modes in thermal quantum field 
theory remained unsettled to date. 

It is the aim of the present article to show that in the case 
of spontaneously broken internal symmetries one can identify 
the Goldstone modes in thermal correlation functions 
by a novel method which reveals their particle nature 
in a clearcut manner. Our approach is based on a general
resolution of thermal correlation functions which seems to be better
adapted to the dissipative effects of a thermal background than mere 
Fourier analysis. This resolution has recently been established 
in the framework of relativistic quantum field theory for
thermal correlation (two--point) functions of arbitrary pairs of
local field operators \cite{BrBu1}. 

Within the present context we are particularly interested in
correlation functions involving the zero component of a conserved
current $j_\mu$ and a field $\phi$,  
\begin{equation} \langle \, j_0 (x) \phi (y) \, \rangle_\beta = 
\lim \frac{1}{Z} \Tr \, e^{- \beta H} \, j_0 (x) \phi (y). \label{1}   
\end{equation}
\noindent Here $\beta > 0$ is the inverse temperature and the right
hand side of relation (\ref{1}) is a reminder of the fact that the correlation
functions are obtained by performing the Gibbs construction in a
finite volume and proceeding to the thermodynamic limit. Assuming that
the state $\langle \, \cdot \, \rangle_\beta$ is invariant under
spacetime translations and that $j_\mu (x)$ and $\phi (y)$ commute at
spacelike distances, $(x-y)^2 < 0$, we can apply the results in
\cite{BrBu1}, giving the following resolution of the correlation
functions (where we have put $y=0$ in order to simplify the notations), 
\begin{equation} \langle \, j_0 (x) \phi (0) \, \rangle_\beta = 
\int_0^\infty dm \, \big( \Dpx \, \pde \Wbo + 
\Dmx \Wbo \big). \label{2} \end{equation}  
Here $(x_0, \bx)$ denote the privileged time and space coordinates of
$x$ in the Lorentz system fixed by the thermal state, 
$\partial_0$ is the time derivative, 
$\Dpmx$ are tempered distributions (depending on the
underlying theory) and 
\begin{equation} \Wbo = (2 \pi )^{-3} \int d^{\, 4} p \, 
\varepsilon ( p_0 ) \delta ( p^2 - m^2 ) (1 - e^{- \beta p_0} )^{-1} 
e^{-ipx} \end{equation} 
\noindent is the two--point function of a free scalar field of mass
$m$ in a thermal equilibrium state at inverse temperature $\beta$. The
corresponding resolution of the commutator function is given by 
\begin{equation} \langle \, [ j_0 (x), \phi (0) ] \, \rangle_\beta = 
\int_0^\infty dm \, \big( \Dpx \, \pde  \Dxm + 
\Dmx \Dxm \big), \label{4} \end{equation}
\noindent where $\Dxm$ denotes the Pauli--Jordan distribution 
\begin{equation} \Dxm = (2 \pi )^{-3} \int d^{\, 4} p \, 
\varepsilon (p_0) \delta (p^2 - m^2) e^{-ipx}.  \end{equation}
\noindent Similar resolutions hold for the time--ordered, advanced and
retarded functions. 

As proved in \cite{BrBu1}, 
relations (\ref{2}) and (\ref{4}) are consequences of the basic principle
of {relativistic causality \/} and provide a natural generalization of
the K\"all\'en--Lehmann representation in the vacuum theory to the
case of thermal equilibrium states. This relationship becomes even more 
obvious if one also takes into account the analyticity properties of 
thermal correlation functions in the 
complex spacetime variables which follow from
the relativistic form of the energy--momentum spectrum in the  
vacuum sector (relativistic KMS--condition \cite{BrBu2}).  
We will make use of this additional structure in the 
subsequent discussion of our results.  

The representations (\ref{2}) and (\ref{4}) show that, in complete analogy to 
the vacuum case, the relativistic thermal correlation functions can be 
resolved 
into a superposition of free field thermal correlation functions of mass 
$m \geq 0$. But whereas the coefficients
$\Dpmx$ are constant 
in $\bx$ in the vacuum theory they exhibit 
in the case of thermal states in general 
a non--trivial $\bx$--dependence, describing the 
dissipative effects of the thermal background on excitations induced by the 
fields. As these effects give rise to the damping of
amplitudes, the coefficients $\Dpmx$ were called {\em damping factors} in 
\cite{BrBu3}. Note that the multiplicative action of the damping factors on  
the free field correlation functions 
amounts to a convolution in momentum space which  
smoothes out in general the discrete mass--shell contributions of stable 
particles, in agreement with the general statements proved in 
\cite{NaReTh}. However, as was shown in \cite{BrBu3}, such 
particles may still be identified 
by discrete mass--contributions appearing in the 
resolutions (\ref{2}) and (\ref{4}) in {configuration space}. 
It is this observation that we intend to use for the identification
of Goldstonean massless particles associated with spontaneous 
symmetry breaking in thermal equilibrium states. 

Before we can apply our representations to the problem of spontaneous
symmetry breaking, we have to comment on the precise mathematical
meaning of relations (\ref{2}) and (\ref{4}), which may not be
completely obvious in view of the singular nature of the
quantities involved. The left hand side of relation (\ref{4}), say,
is defined in the sense of tempered distributions, i.e., becomes
meaningful if integrated with a test function. It suffices for our
purposes to consider test functions of the form $f (x_0) g( \bx )$,
where $f, g$ have compact support. For the specification of the right
hand side of relation (\ref{4}) we make use of the fact that 
\begin{equation} \hpmx \doteq \int dx_0 \, f(x_0) \, \partial^{(\pm)} 
\Dxm, \label{6} \end{equation} 
\noindent (where $\partial^{(+)} \doteq \pde, 
\partial^{(-)} \doteq 1)$ are test functions in the variables $\bx, m$
for any choice of the test function $f$. The rigorous version of
relation (\ref{4}) can then be presented in the form 
\begin{eqnarray} \lefteqn{\int d^{\, 4} x \, f (x_0) g( \bx )
\, \langle \, [ j_0 (x), \phi (0) ] \, \rangle_\beta =}  \label{7} \\ 
& & = \int d^{\, 3} \bx \int dm \, \big( \Dpx g( \bx ) \hpx + 
\Dmx g( \bx ) \hmx \big), \nonumber 
\end{eqnarray} 
\noindent where the right hand side of (\ref{7}) is meaningful since
$\Dpmx$ are tempered distributions with support in $\RR^3 \times
\RR_+$. In a similar fashion one can give a precise meaning to relation
(\ref{2}). 

Let us now turn to the discussion of the symmetry transformation
induced by the current $j_\mu$. The charge operator corresponding to
this current (if it exists) is defined as a suitable limit of the operators 
\begin{equation} Q_R \doteq \int d^{\, 4} x \, f(x_0) g(
\bx / R) \, j_0 (x) \label{8} 
\end{equation} 
for $R \rightarrow \infty$, where $g$ is any test function which is
equal to $1$ in the unit ball about the origin of configuration
space and $f$ is usually normalized according to $\intf = 1$. (In
order to display the role of $f$ in the subsequent argument we do not
impose here the latter condition, however.) The operators $Q_R$ are thus 
the appropriately regularized charge operators for 
finite spatial volume. 

It follows from the spacelike commutativity of $j_\mu (x), \phi (y)$
and current conservation that for sufficiently large $R$ (depending on
the choice of the function $f$) there holds 
\begin{equation} \langle \, [ Q_R, \phi (0) ] \, \rangle_\beta = 
q \, \intf, \label{9} \end{equation} 
\noindent where $q$ is some constant which does not depend on the choice of
$f,g$ within the above limitations. This result can be established by
the same arguments as in the vacuum theory \cite{Yy}, so we
do not need to reproduce it here. The symmetry corresponding to the
current $j_\mu$ is said to be spontaneously broken in the
state $\langle \, \cdot \, \rangle_\beta$, if $q \neq 0$ for some
field $\phi$. Then the limit of the operators $Q_R$ does not
define the generator of a unitary group which induces the symmetry
transformation and leaves the state $\langle \, \cdot \,
\rangle_\beta$ invariant.

As is well known, Eqs.\ (\ref{8}) and (\ref{9}) imply that the Fourier 
transform of the commutator function 
$\langle \, [ j_0 (x), \phi (0) ] \, \rangle_\beta$ 
is equal to $(q / 2 \pi) \, \delta (p_0)$ for spatial momentum $\bp = 0$.  
But, in contrast to the vacuum case, the occurrence of a 
$\delta$--singularity on the full light cone $p^2 = 0$ is not implied 
by this fact. So there is {\em a priori\/} no basis for the 
conventional momentum-space
interpretation of these zero--energy modes as relativistic particles. 
As already mentioned, 
there are examples just outside the present framework 
(based on currents $j_\mu (x)$ which depend explicitly on 
the spacetime coordinates $x$) where these modes 
cannot be associated with a particle \cite{Oj}. 
However, for the local covariant  
currents considered here, we are able to show that if relation 
(\ref{9}) holds for some $q \neq 0$, there 
appear (in a generic way) certain specific discrete contributions in 
the damping factor $\Dpx$ in our representation 
(\ref{2}) which can be attributed to a Goldstone boson.
These particles can thus be ``unmasked'' in configuration space. 

Turning to the proof of this statement we first note that for odd 
test functions $f$ the right hand side of relation (\ref{9}) vanishes. 
So we may restrict attention to even
$f$ and proceed, by combining relations (\ref{7}) to (\ref{9}), to
\begin{equation} \int d^{\, 3} \bx \int dm \, \Dpx \, g ( \bx / R ) \hpx
= q \, \intf \label{10} \end{equation} 
\noindent which holds for sufficiently large $R$. Since $g(0) = 1$, the
sequence of test functions $ g ( \bx / R ) \hpx $ converges for $R
\rightarrow \infty$ to $ \hpx$ (in the appropriate topology),  
hence equation (\ref{10}) becomes in this limit 
\begin{equation} \int d^{\, 3} \bx \int dm \, \Dpx \hpx = q \, \intf. 
\label{11}  \end{equation}
\noindent The constraints imposed on $\Dpx$ by the latter equation are
more transparent in momentum space. There one gets, 
bearing in mind the definition (\ref{6}) of $\hpx$, 
\begin{equation} \int d^{\, 3} \bp \int dm \, \Dpp \widetilde{f} 
( \sqrt{ \bp^2 + m^2 }) = iq \, (2 \pi )^{3/2} \widetilde{f} (0), 
\label{12} \end{equation} 
\noindent where $\widetilde{f} ( p_0) $ denotes the Fourier transform of
$f(x_0)$ and $\Dpp$ the (partial) Fourier transform of $\Dpx$ with
respect to the spatial variables $\bx$. 

Relation (\ref{12}) holds for any choice of the (even) test
function $f$. The corresponding functions $\widetilde{f} ( \sqrt{
\bp^2 + m^2 })$ on the space $\RR^4$ spanned by $\bp, m$ exhaust the
set of all spherically symmetric test functions on this space. Hence we
infer from (\ref{12}) that in the ({unique\/}) decomposition of
the distribution $\Dpp$ into its spherically symmetric part 
in $(\bp,m)$--space and the
remainder, the symmetric part consists of a multiple of the
$\delta$--function, 
\begin{equation} \Dpp = iq \, (2 \pi )^{3/2} \delta ( \bp ) \delta (m)
+ {\Rpp}, \label{13} \end{equation} 
\noindent and going back to configuration space we arrive at 
\begin{equation} \Dpx = iq \, \delta (m) + \Rpx. \label{14}
\end{equation} 

This relation provides some evidence to the effect that in $\Dpx$ 
there appears generically a discrete zero mass contribution 
if the symmetry is spontaneously broken. Yet this idea     
requires some further analysis since the remainder in (\ref{14})
could contain additional singular terms which completely screen the 
effects of the delta function; in other words, 
relation (\ref{14}) does not establish a decomposition
of the damping factor into contributions of different degree 
of singularity. In order to clarify this point we consider in the 
following the physically interesting case where \, $dm\,\Dpx $  \, 
is a (complex) measure in $m$ which is regular 
in $\bx$. The latter property follows from the relativistic 
KMS--condition \cite{BrBu1} and the former one may be expected 
to hold quite generally, similarly to 
the case of the ``weight functions'' in the K\"all\'en--Lehmann 
representation of vacuum correlation functions, cf.\ \cite{BrBu3}. 
We recall that according to standard theorems on the decompositon 
of measures the statement that a measure has a  
discrete contribution has a clearcut mathematical meaning.   
Moreover, discrete contributions are the most 
prominent singularities which can occur. 

In order to establish rigorously from equation (\ref{12}) 
that such a discrete zero mass contribution is present in $\Dpx$
in generic cases, let us make  
the additional physically motivated 
assumption that the damping factor decreases for
increasing $| \bx |$. More precisely, presenting $\Dpx $ as the 
Radon--Nikodym derivative of a piecewise continuous, bounded function, 
\begin{equation} \Dpx = \partial_m \Cpx, \label{15} \end{equation}
let us assume that $| \Cpx |$ is monotonically decreasing in 
$| \bx |$ for fixed $\bx / |\bx|$ and $m$. Such a behavior 
of the damping factors may be expected 
in the presence of dissipative effects of a 
spatially homogeneous thermal background \cite{BrBu3}. 

By writing equation (\ref{12}) with the choice of Gaussian 
functions $\widetilde{f} (p_0) = e^{-\lambda^2 p_0^2 /2 }$, 
one obtains after a straightforward
computation  
\begin{eqnarray} 
iq \, (2 \pi )^{3/2} & = & \lim_{\lambda \rightarrow \infty}
\int d^{\, 3} \bp \int dm \, \Dpp  \, \label{16}
e^{-\lambda^2 ( \sbp^2 + m^2 )/2} \nonumber \\
& = & \lim_{\lambda \rightarrow \infty} 
\int d^{\, 3} \bx \int dm \, m \, \Cplx  \, e^{- ( \sbx^2 + m^2 )/2}.
\end{eqnarray}
It therefore follows  
(in view of the dominated convergence theorem,    
which governs the interchange of limits and integrations) that 
the function $\Cpx$ cannot exhibit a 
``trivial scaling behavior'' of the form
\, $\lim_{\lambda \rightarrow \infty} \Cplx \, = \, 0$ \, 
for all $\bx$ and fixed $m > 0$  
if $q \neq 0$. On the other hand, the monotonicity
properties of $\Cpx$ imply that for $\lambda \geq 1$ 
there holds
$ |C^{(+)}_\beta ({\bx}, \lambda^{-1} m)| \geq 
|C^{(+)}_\beta (\lambda {\bx}, \lambda^{-1} m)| $.
Proceeding to the limit $\lambda \rightarrow \infty$ we 
therefore conclude that 
\begin{equation} 
\lim_{m \,\downarrow \ \! 0} \, C^{(+)}_\beta ({\bx}, m) 
\not\equiv 0. \label{17}
\end{equation}
Since $C^{(+)}_\beta ({\bx}, m)$ 
vanishes for negative $m$ this shows that 
the latter function is discontinuous 
in $m$ at $m=0$ and consequently 
(in view of (\ref{15})) $\Dpx$ has a discrete zero mass
contribution of the form $C_\beta ({\bx}) \, \delta (m)$, where 
$C_\beta ({\bx})$ denotes the limit in (\ref{17}). 
(It can be seen that the same conclusion also holds under considerably
weaker assumptions.)  

There is another interesting consequence of relation (\ref{16}) 
with regard to 
the spatial behavior of the damping factors. Because of the 
dissipative effects already mentioned one expects that the 
function $\Cpx$ tends to 0 for large $|\bx|$ for almost all
$m$. We now claim that this decay has to be slow in the presence of 
spontaneous symmetry breaking. For if there holds 
for large $|\bx|$ and small $m$
\begin{equation}
| m^\delta \, \Cpx | \leq \mbox{\em const\/} \, 
|\bx|^{-\varepsilon}    \label{18}
\end{equation}
for some $\varepsilon > \delta  \geq 0$, 
it follows from (\ref{16}) that $q = 0$. 
This result is in accord with the physical picture that 
sufficiently strong dissipative  
effects destroy the long range order in thermal  
states and thereby lead to a restoration of symmetries. 

Note however that the bound in  
(\ref{18}) does not exclude the presence of a discontinuity of 
$\Cpx$  at $m = 0$. This shows that, in contrast to the 
vacuum case, there may appear
discrete massless contributions in the damping factors of 
current--field correlation functions without spontaneous breakdown 
of symmetries and this occurs as an effect of the strength of the damping. 
This phenomenon is consistent with the following scenario which 
is of interest from a physical viewpoint:
if some spontaneously broken symmetry, which is accompanied by a 
Goldstone particle, is restored at high temperatures, this particle 
need not cease to exist. From a more technical viewpoint,
based on the general form (\ref{15}) of our damping factor 
$\Dpx$, what matters for the breakdown of 
symmetries is the scaling behavior of $\Cpx$ whose non--vanishing
character can be considered as a criterion for symmetry breaking.  

To summarize, we have seen in the preceding analysis that
in the case of spontaneously broken internal symmetries there appear  
generically in the representation (\ref{2}) 
of current--field correlation functions damping 
factors $\Dpx$ which 
\begin{itemize}
\item[(a)] {contain a discrete (in the sense of measures) 
zero--mass contribution and}
\item[(b)] {are slowly decreasing in $| \bx |$ for small 
values of $m$.}
\end{itemize}
So these damping factors coincide {locally\/} with 
the K\"all\'en--Lehmann weights appearing in the case of 
spontaneous symmetry breaking in the vacuum sector.  
Hence, adopting the arguments in  
\cite{BrBu3}, the particle nature of the Goldstone
modes in thermal equilibrium states may be regarded as settled. 

It is easily seen in examples that the above mentioned properties 
of the damping factors do not imply that there holds a sharp
energy--momentum dispersion law for the Goldstone particles. 
This fact corroborates the statement made in \cite{BrBu3} 
that the K\"all\'en--Lehmann type 
representation (\ref{2}) is better suited than Fourier 
analysis to uncover the particle aspects of thermal equilibrium 
states. The representation is also useful for the analysis of the 
more subtle momentum space properties of thermal correlation functions,  
where one has to employ techniques of 
complex analysis. We intend to return to this interesting 
issue elsewhere.\\[2mm]

\noindent {\Large\bf Acknowledgements} \\[1mm]
The authors are grateful for financial support by the Franco--German
science cooperation PROCOPE.


\begin{thebibliography}{9}
\vspace*{-2mm}
\bibitem{Ka} 
J.I.\ Kapusta, 
{\em Finite--Temperature Field Theory\/} (Cambridge University Press, 
Cambridge 1993). 
 
\bibitem{NaReTh}
H.\ Narnhofer, M.\ Requardt and W.\ Thirring, 
Commun.\ Math.\ Phys.\ 92, 247 (1983). 

\bibitem{Um} 
H.\ Umezawa, 
{\em Advanced Field Theory\/} 
(American Institute of Physics, New York 1993). 

\bibitem{Oj}
I.\ Ojima, Lecture Notes in Physics, No.\ 176 (Springer, Berlin 1983),
pp.\ 161--165; \, Lett.\ Math.\ Phys.\ 11, 73 (1986).   

\bibitem{BrBu1}
J.\ Bros and D.\ Buchholz,
(to be published);  cf.\ also 
Annales Inst.\ H.\ Poincar\'e 64, 495 (1996).                    

\bibitem{BrBu2}
J.\ Bros and D.\ Buchholz, 
Nucl.\ Phys.\  B429, 291 (1994). 

\bibitem{BrBu3}
J.\ Bros and D.\ Buchholz,
Z.\ Phys.\ C.\ Particles and Fields 55, 509 (1992). 

\bibitem{Yy}
J.A.\ Swieca,
in: Carg\'ese Lectures, Vol.\ 4, 
edited by D.\ Kastler (Gordon and Breach, New York 1969). 

\end{thebibliography}
\end{document}